\documentclass[
12pt,
showkeys,
 amsmath,amssymb,
aps,
]{revtex4-2}

\usepackage{graphicx}
\usepackage{dcolumn}
\usepackage{xcolor}
\usepackage{bm}

\begin{document}


\title{Discovery of magnetic phase transitions in heavy-fermion superconductor CeRh$_2$As$_2$}

\author{Grzegorz Chajewski}
 \email{g.chajewski@intibs.pl}
\author{Dariusz Kaczorowski}
\email{d.kaczorowski@intibs.pl}%
\affiliation{Institute of Low Temperature and Structure Research, Polish Academy of Sciences, Ok\'olna 2, 50-422 Wroc{\l}aw, Poland }%



\begin{abstract}
We report on the specific heat studies performed on a new generation of CeRh$_2$As$_2$ single crystals. Superior quality of the samples and dedicated experimental protocol allowed us to observe an antiferromagnetic-like behavior in the normal state and to detect the first-order phase transition of magnetic origin within the superconducting state of the compound. Although in the available literature the physical behavior of CeRh$_2$As$_2$ is most often described with the use of quadrupole density wave scenario, we propose an alternative explanation using analogies to antiferromagnetic heavy-fermion superconductors  CeRhIn$_5$ and Ce$_2$RhIn$_8$.
\end{abstract}

\maketitle
Heavy-fermion superconductivity is one of the most intriguing phenomena of modern solid-state physics. Its full understanding seems to remain elusive at the current stage, and continuous studies bring new questions and challenges. The recently discovered heavy-fermion superconductor CeRh$_2$As$_2$ \cite{Khim2021} focused enormous attention of solid-state community due to its unusual features. One of these properties is the phase transition within the superconducting state, which occurs upon applying magnetic field along the \textit{c}-axis and leads to an extraordinary violation of the Pauli-Clogston limit. It also makes CeRh$_2$As$_2$ one of the only few known multi-phase heavy-fermion superconductors (together with e.g. UTe$_2$ \cite{Braithwaite2019,Aoki2020} and UPt$_3$ \cite{Fisher1989,Bruls1990,Adenwalla1990}). The second interesting feature of the compound is the enigmatic $T_0$-anomaly, observed at temperature slightly above the superconducting phase transition. Although numerous experimental investigations \cite{Hafner2022, Landaeta2022, Onishi2022, Kimura2021, Kibune2022, Kitagawa2022, Siddiquee2022, Mishra2022, Semeniuk2023} supported by various theoretical studies \cite{Schertenleib2021, Mockli2021a, Mockli2021b, Ptok2021, Cavanagh2022, Nogaki2021, Hazra2022, Nogaki2022, Machida2022} have been performed up to date, there is still no decisive explanation for the nature of both these peculiarities. It has been proposed that CeRh$_2$As$_2$ hosts a quadrupole density wave (QDW) instability \cite{Hafner2022} and the anomaly at $T_0$ is a signature of the entrance to this unique non-magnetic ordered state. Interestingly, Machida recently proposed \cite{Machida2022} the antiferromagnetic (AFM) ordering picture for an explanation of the anomaly in the field dependence of magnetic susceptibility, $\chi (H)$, of CeRh$_2$As$_2$, which marks the transition between its two superconducting states. Here we show that the same scenario can also describe the behavior of the specific heat data, $C_{\rm p}(T)$, of this compound without including the QDW concept.

The insufficient homogeneity of the samples reported so far, responsible for significant broadening of thermodynamic features related to the phase transitions, is an essential problem in the proper understanding of the physical behavior of CeRh$_2$As$_2$. 
In general, high-quality samples are crucial for detailed investigation of all low-temperature unconventional superconductors. As recent studies on UTe$_2$ \cite{Ran2019, Matsumara2023}, Sr$_2$RuO$_4$ \cite{Steppke2017, Pustogow2019}, and
YbRh$_2$Si$_2$ \cite{Schuberth2016, Nguyen2021} have shown, the superior quality of samples, either solely, or combined with other developments (like strain tuning of the critical temperature or precise temperature control of the experimental setup) provided important insights into the fascinating nature of these materials.
Recently Semeniuk \textit{et al}. reported on a new batch of single crystals of CeRh$_2$As$_2$ \cite{Semeniuk2023}, the quality of which was claimed to be the highest among those investigated before, based on a few parameters derived from the thermodynamic and electrical transport data. Analysis of the very same set of parameters for crystals synthesized and studied in the present work (see Supplementary Information \cite{supplement}\nocite{Chajewski2023, Hwang1998, Scheie2018, Lengyel2008}) indicated that they are of even higher quality. The already reported anomalies in CeRh$_2$As$_2$ have been reproduced in our measurements, however, they are sharper and clearer. Most importantly, we have also observed a new feature in $C_{\rm p}(T)$, which was not found in the previous studies possibly because of inhomogeneity of samples and/or different experimental protocols.


\begin{figure}[htbp]%
\centering
\includegraphics[width=1\linewidth]{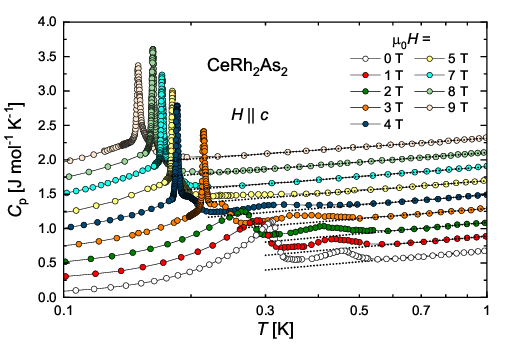}
\caption{Specific heat of CeRh$_2$As$_2$ in magnetic fields applied along the \textit{c}-axis. Straight dashed lines emphasize a logarithmic behavior. For clarity, the consecutive curves were shifted upwards by 0.2~Jmol$^{-1}$K$^{-1}$. } \label{fig1}
\end{figure}

Fig.~\ref{fig1} presents low-temperature $C_{\rm p}(T)$ data collected in several different magnetic fields applied along the \textit{c}-axis for the new generation of CeRh$_2$As$_2$ single crystals. As can be easily noticed, the zero-field curve exhibits two distinct anomalies. A smaller one, at $T_0$~=~0.50~K, has been proposed to be the manifestation of a QDW \cite{Hafner2022, Semeniuk2023}. In turn, the larger one, at $T_{\rm sc}$~=~0.32~K, is related to the superconducting phase transition. The determined temperatures are distinctly higher than temperatures originally reported by Khim \textit{et al}. \cite{Khim2021} and comparable with the ones revealed recently \cite{Semeniuk2023}. Above $T_0$, up to about $T$~=~3~K for the zero-field data (for the extended data range see Supplementary Information \cite{supplement}), the normal state $C_{\rm p}(T)$ curve can be very well described by logarithmic temperature dependence $A\ln(T/T^*)$, where $A$ and $T^*$ are parameters of the fit. The same relation can be also fitted at least up to 1~K to the corresponding experimental data acquired with applied magnetic fields (see the straight dashed lines in Fig.~\ref{fig1}a).  While the  parameter $A$ with increasing magnetic field changes only slightly (from $0.26$ ~Jmol$^{-1}$K$^{-1}$ for 0~T to $0.21$~Jmol$^{-1}$K$^{-1}$ for 9~T), the corresponding change of $T^*$ is much more pronounced (from 53~mK to 23~mK). It should also be noted, that with the shift of $T_0$ towards lower temperatures, the low-temperature limit of the applicability of this relation lowers accordingly. Assuming that the phonon contribution to the specific heat of CeRh$_2$As$_2$ is similar to that determined for its non-magnetic counterpart LaRh$_2$As$_2$ ($\beta$~=~0.34~mJmol$^{-1}$K$^{-4}$ \cite{Landaeta2022La}), it can be safely stated that below 1~K the lattice contribution to $C_{\rm p}$ is negligibly small compared to the measured specific heat in this temperature range. Hence, the logarithmic term ${\sim}\ln T$ provides a good description of the sum of electronic and magnetic contributions to $C_{\rm p}(T)$ of CeRh$_2$As$_2$. This is somehow surprising, since in Ce-based compounds for the low-temperature non-Fermi-liquid  behavior one usually expects the ${\sim}T\ln T$ dependence of the specific heat. Interestingly, based on the high-pressure electronic specific heat data of Ce$_2$RhIn$_8$ \cite{Lengyel2004} we found that, regardless of the applied pressure, the high-temperature tail of the antiferromagnetic anomaly in Ce$_2$RhIn$_8$ in the temperature range from about 3 to 7~K can be very well described by the same logarithmic function $A\ln(T/T^*)$, as was used to describe the normal state specific heat of CeRh$_2$As$_2$.

As can be seen in Fig.~\ref{fig1}, with increasing strength of the magnetic field, $T_0$-anomaly broadens and systematically shifts towards lower temperatures, which is a feature typical for antiferromagnets. It can also be noticed that because with increasing magnetic field the anomaly at $T_{\rm sc}$ shifts towards lower temperatures slower compared to the $T_0$-anomaly, with rising field strength they get closer and closer to one another and above about 5~T only a single distinct anomaly is visible.  

\begin{figure}[htbp]%
\centering
\includegraphics[width=1\linewidth]{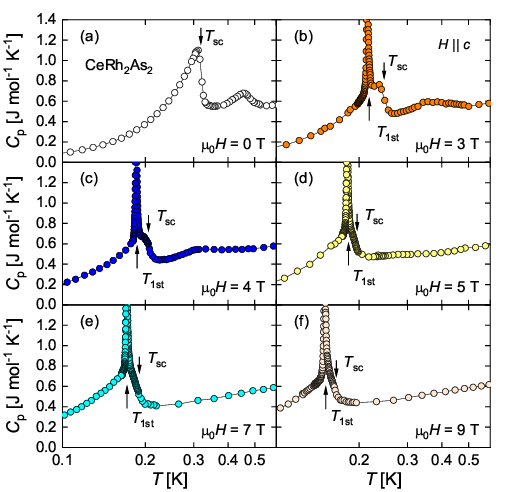}
\caption{Evolution of the anomalies in the specific heat of CeRh$_2$As$_2$ in magnetic fields of (a) 0~T, (b) 3~T, (c) 4~T, (d) 5~T, (e) 7~T, and (f) 9~T applied along the \textit{c}-axis. The arrows mark the phase transitions at $T_{\rm sc}$ and $T_{\rm 1st}$.} \label{fig2}
\end{figure}

Importantly, by performing specific heat measurements with the use of the long heat pulse method \cite{Scheie2018}, in magnetic fields higher than about 3~T, we were also able to detect a first-order phase transition which occurs within the superconducting state of the compound. No such a transition was clearly visible in $C_{\rm p}(T)$ curves neither for zero magnetic field (see Fig.~\ref{fig2}a) nor any other magnetic field smaller than about 3~T. Taking a closer look at the specific heat curves measured in different magnetic fields, one can observe that in 3~T the superconducting phase transition at $T_{\rm sc}$ and the first-order transition at $T_{\rm 1st}$ are clearly separated, as shown in Fig.~\ref{fig2}b. In higher fields they approach each other (see Fig.~\ref{fig2}c) and above about 5~T (Fig.~\ref{fig2}d) the first-order-type anomaly follows the superconducting one as if they are coupled, shifting together towards lower temperatures with further increase of magnetic field (Fig.~\ref{fig2}e and \ref{fig2}f). Because to the best of our knowledge no such a behavior has been reported up to date for a purely superconducting system, we suppose that the first-order transition is closely related to the $T_0$-anomaly. 
In turn, very similar field-induced phase transitions have already been observed e.g. for antiferromagnetic superconductors CeRhIn$_5$ and  Ce$_2$RhIn$_8$ \cite{Cornelius2001, Lengyel2008}. For both compounds, the first-order phase transition has been proven to occur due to the reconstruction of their antiferromagnetic structures \cite{Bao2001, Raymond2007}. Therefore, the presence of the phase transition at $T_{\rm 1st}$ can be considered as a strong support of the AFM scenario, which presumes that the $T_0$-anomaly in CeRh$_2$As$_2$ is a manifestation of some kind of long-range magnetic ordering. Such previously unobserved behavior may also provide an alternative explanation of the mechanisms governing the change of the superconducting state of CeRh$_2$As$_2$, to be discussed later. More precise studies of the evolution of the first-order phase transition were performed using a set of various specific heat measurement approaches, and the details are described in Supplementary Information \cite{supplement}.

\begin{figure}[htbp]%
\centering
\includegraphics[width=1\linewidth]{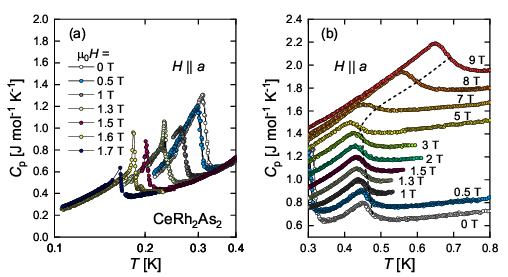}
\caption{Specific heat of CeRh$_2$As$_2$ near the (a) superconducting transition and (b) magnetic transition in magnetic fields applied along the \textit{a}-axis. The dashed curve tracks the position of $T_0$.  For clarity, the consecutive curves were shifted upwards by 0.1 Jmol$^{-1}$K$^{-1}$. } \label{fig3}
\end{figure}

Fig.~\ref{fig3} displays low-temperature specific heat data of CeRh$_2$As$_2$ collected for $H \parallel a$. As can be inferred from Fig.~\ref{fig3}a, in this configuration, the behavior of the superconductivity-related anomaly is initially typical for most superconductors. With increasing magnetic field, the anomaly at $T_{\rm sc}$ shifts to lower temperatures and its magnitude diminishes. Interestingly, in a field of about 1.3~T, the  peak in $C_{\rm p}(T)$ becomes distinctly sharper, hence suggesting a change in character of the phase transition from the second to first-order. A similar effect was observed in e.g. CeCoIn$_5$ \cite{Bianchi2003} and Ce$_2$PdIn$_8$ \cite{Tokiwa2011} and was ascribed to possible emergence of low-temperature high-field FFLO (Fulde-Ferrell-Larkin-Ovchinnikov) state \cite{Bianchi2003, Dong2011}. It is a tempting conjecture that the observed behavior is a manifestation of FFLO state formation in CeRh$_2$As$_2$, however, this hypothesis needs solid experimental verification.

The evolution of the $T_{0}$-anomaly in $H \parallel a$ is presented in Fig.~\ref{fig3}b. Since measurements for this field direction were performed on a different piece of single crystal, one can notice slight differences in the shape of $T_0$-anomaly. Compared to that used for $H \parallel c$ measurements, it is sharper and little higher, which suggests even better homogeneity of this part of the crystal. As a consequence of reduced broadening of the anomaly, the transition temperature $T_0$~=~0.48~K is somewhat lower. As can be noticed, for fields lower than about 5~T, with increasing field strength $T_0$ slightly decreases, as expected for antiferromagnets. Above 5~T, this trend changes and further increase of the magnetic field results in a distinct shift of the anomaly towards higher temperatures. At the same time the $\lambda$-shaped anomaly becomes sharper. This strange feature of CeRh$_2$As$_2$ has not been reported in the previous studies.

\begin{figure}[htbp]%
\centering
\includegraphics[width=1\linewidth]{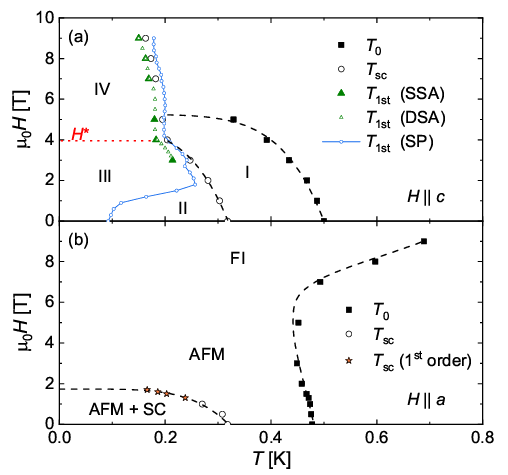}
\caption{\textit{H-T} phase diagrams of CeRh$_2$As$_2$ constructed from the specific heat data collected in magnetic field applied along (a) the \textit{c}-axis and (b) the \textit{a}-axis. Dashed lines are the guides for the eye. The red dotted line marks the approximate position of the  $H^*$ anomaly taken from the magnetic susceptibility data \cite{Khim2021}. For the explanation of abbreviations see the main text.} \label{fig4}
\end{figure}

Based on our specific heat data, we constructed the phase diagrams of CeRh$_2$As$_2$, which are presented in Fig.~\ref{fig4}. $T_{sc}$ and $T_{0}$  were determined by equal entropy construction. In turn, the values of $T_{\rm 1st}$ for data obtained by both single-slope (SSA) and dual-slope analysis (DSA) (details on both analysis methods can be found in e.g. Ref.~\cite{HCPPMS}) were taken from the positions of the maxima of the anomalies.  In the diagram for $H \parallel c$, we also included the field variation of $T_{\rm 1st}$ measured by the sample platform thermometer for a base temperature of 60~mK, here denoted as $T_{\rm 1st}$ (SP) (see Supplementary Information \cite{supplement} for details). It should be noted that derived in different ways values of $T_{\rm 1st}$ are in very good agreement with each other and a little shift of the $T_{\rm 1st}$ (SP) line towards higher temperatures results from the different definition of the transition temperature used in this case.
The overall shape of the diagrams, except for the $T_{\rm 1st}$ line, that is shown here for the first time, is consistent with the previously reported ones \cite{Khim2021, Mishra2022, Semeniuk2023}. The newly-discovered $T_{\rm 1st}$ boundary appears only for $H \parallel c$, and in small magnetic fields exists well below $T_{\rm sc}$. However, with increasing magnetic field strength  its position gets closer to $T_{\rm sc}$, still staying within the superconducting state. In $H \approx$ 4~T (being very close to $H^*$ determined from the magnetic susceptibility measurements \cite{Khim2021}), the anomalies at $T_{\rm sc}$ and $T_{\rm 1st}$  merge, and in higher fields, they follow the same path. 

In view of the presented experimental results, we describe CeRh$_2$As$_2$ as an antiferromagnetic superconductor, which in magnetic fields applied along the \textit{c}-axis undergoes metamagnetic-like phase transition at $T_{\rm 1st}$, related to some change in the spin structure. 
 Within this scenario, several different regions of the phase diagram of CeRh$_2$As$_2$ can be distinguished for $H \parallel c$ (see Fig.~\ref{fig4}a). In region I, the compound is likely an antiferromagnet with the magnetic moments aligned most probably along the \textit{c}-axis. One should note that this picture agrees with one of the proposed magnetic structures describing the AFM state detected by the nuclear quadrupole resonance (NQR) measurements \cite{Kibune2022}. However, by analogy to Ce$_m$RhIn$_{3m+2}$ compounds, one can expect for CeRh$_2$As$_2$ a more complex magnetic structure (such as incommensurate spin density wave) rather than simple AFM phase. At lower temperatures (region II), antiferromagnetism and superconductivity coexist in the material, similarly as was observed for CeRhIn$_5$ and Ce$_2$RhIn$_8$ under external pressure \cite{Fisher2003, Knebel2004, Knebel2006, Knebel2011, Nicklas2003}. The transition from region II to region III may signal the transformation of the magnetic structure of CeRh$_2$As$_2$, possibly into a commensurate colinear AFM one. Consequently, in region III, rising magnetic field brings about gradual canting of the moments initially oriented antiparallel to the field direction. In a field $H^*$, a metamagnetic transition occurs, possibly of a spin-flop character, and in fields $H > H^*$ (region IV), a growing ferromagnetic component develops. Thus, in this region the superconducting state coexists with the field-induced ferromagnetic-like order instead of the AFM one. As a support for this hypothesis, one may consider the S-shape of the temperature dependence of $H_{\rm {c2}}(T)$, determined for the high-field superconducting phase, that resembles those observed for ferromagnetic superconductors UGe$_2$ and UCoGe (see e.g. ref. \cite{Aoki2019}). Moreover, our scenario is compatible with the theoretical description proposed recently by Machida \cite{Machida2022}. 
 
 Although the aforementioned NQR experiments \cite{Kibune2022}, as well as recent nuclear magnetic resonance (NMR) studies \cite{Ogata2023} detected the antiferromagnetic ordering in CeRh$_2$As$_2$, it needs to be emphasized that according to those studies, the AFM order emerges within the low-field superconducting state of the compound, i.e. well below the $T_0$ temperature. This finding clearly differs from our interpretation of the $T_0$ as the temperature of transition to the ordered magnetic state. However, we also notice some similarity. In our scenario, for 3~T $< H < H^*$, the commensurate AFM state seems to appear at temperature very close to that of the emergence of AFM order in the NQR and NMR studies. For lower fields, due to different measurement conditions (in NQR and NMR measurements external magnetic field and temperature are kept stable, while in our measurement of $T_{\rm 1st}$ (SP) only the magnetic field is static, but temperature changes dynamically), it is hard to reliably compare our results to those reported before. Nevertheless, the most striking difference is observed for $H > H^*$. In this region, our results show that $T_{\rm 1st}$ merges with $T_{\rm sc}$, hence magnetic ordering appears in the material just after (or together with) the superconductivity emergence. Contrarily, according to the results of NMR measurements \cite{Ogata2023}, antiferromagnetism  is no longer observed upon transition to the high-field superconducting phase. This leads to the question, what is the origin of $T_{\rm 1st}$ when $H > H^*$, if no magnetism is present in the material in this region, as suggested by NMR studies.

 The phase diagram constructed for $H \parallel a$ is much simpler (see Fig.~\ref{fig4}b). It includes the superconducting phase (SC) which exists within the AFM state and two regions of magnetism of a different character. The antiferromagnetic-like behavior is observed in fields up to about 6~T. In higher fields, most probably due to a field-induced spin structure reconstruction (FI region), the anomaly in $C_{\rm p}(T)$ shifts towards higher temperatures, typically for ferromagnets. This behavior distinctly differs from that observed for $H \parallel c$. However, one should notice that in our scenario, when a magnetic field weaker than about 6~T is applied along the $a$-axis, the orientation of magnetic moments with respect to the external magnetic field might resemble their orientation above spin-flop transition (above 4~T) while the magnetic field is applied along the $c$-axis. In both these cases, the transition temperature slowly shifts towards lower temperatures with increasing field strength. Hence, it is possible that for $H \parallel c$, the field used in our experiments was not strong enough to polarize magnetic moments into a ferromagnetic-like state. One should also take into account the anisotropy of the system as a possible reason for this behavior. To clarify the actual nature of the magnetism and spin structures in CeRh$_2$As$_2$, further meticulous investigation is necessary.

Based on our interpretation of the collected specific heat data, the $T_0$-anomaly is ascribed to AFM ordering, despite so far no distinct signature of the phase transition at $T_0$ was found in the magnetic susceptibility data \cite{Khim2021}. This finding resembles e.g. the case of pressurized EuTe$_2$ \cite{Yang2022}, where in pressures higher than about 10~GPa AFM order manifests itself in $C_{\rm p}(T)$ as a clear $\lambda$-shaped peak at $T_{\rm N}$, but there is no related anomaly in $\chi(T)$. Another explanation of the lack of any clear feature in the magnetic data at $T_0$ may refer to insufficient quality of the first generation of CeRh$_2$As$_2$ single crystals \cite{Khim2021}, for which even the specific heat anomaly at $T_0$ was hardly visible and the anomaly in $\chi(T)$ could possibly be overlooked. Perhaps, a clear magnetic fingerprint of AFM order could be detected in the newly grown samples. Unfortunately, we are not capable to measure magnetic susceptibility on our single crystals at such low temperatures in order to verify this hypothesis.

Another issue is related to the magnetic entropy released by $T_{0}$ in CeRh$_2$As$_2$, which for our single crystals equals about 12\% of $R \ln 2$. It is only a small fraction of the value expected for dublet ground state. However, for example, in pressurized AFM superconductor Ce$_2$RhIn$_8$, similar reduction of entropy at $T_{\rm N}$ ($0.1R\ln 2$ in $p \approx 1.65$~GPa), accompanied with a strong suppression of the peak in $C_{\rm p}(T)$, was observed \cite{Lengyel2008}.


Our detailed specific heat studies of newly grown CeRh$_2$As$_2$ single crystals allowed us to accurately trace the magnetic field evolution of the $T_0$-anomaly, which led us to conclusion on its antiferromagnetic-like character. To describe the magnetically ordered state, we used the scenario proposed recently for the explanation of the magnetic susceptibility features of the compound. We proposed that AFM order in CeRh$_2$As$_2$ may appear above the onset of superconductivity, i.e., as it happens in every other known magnetic heavy-fermion superconductor. This scenario allowed us to consistently describe all our experimental data without resorting to the concept of QDW formation at $T_0$. It should be noted that the latter picture implies that CeRh$_2$As$_2$ showing the onset of AFM inside the superconducting dome would be unique amidst heavy-fermion superconductors. We also pointed out a few similarities of CeRh$_2$As$_2$ to heavy-fermion antiferromagnetic superconductors CeRhIn$_5$ and Ce$_2$RhIn$_8$.
Moreover, due to the high quality of our single crystals, we revealed the first-order phase transition at $T_{\rm 1st}$, which was not detected in any previous investigation. We ascribed it to the transformation of the magnetic structure of the compound, tentatively from incommensurate spin density wave to commensurate linear AFM one. In our opinion, another change in the magnetic structure, observed in $H^* \approx$ 4~T, may play a role of a driving force for the change of the superconducting state of CeRh$_2$As$_2$.

\newpage

\begin{center}
\textbf{\large Supplementary Information}
\end{center}
\setcounter{equation}{0}
\setcounter{figure}{0}
\setcounter{table}{0}
\setcounter{section}{0}

\maketitle
\renewcommand{\bibnumfmt}[1]{[S#1]}
\renewcommand{\citenumfont}[1]{S#1}
\renewcommand{\thefigure}{S\arabic{figure}}

\section{Experimental methods}\label{sec4}

\textit{\textbf {Single crystal growth}}

For the purpose of this study a new generation of single crystalline samples of CeRh$_2$As$_2$ has been grown using an appropriately adapted horizontal flux growth technique. The detailed synthesis procedure can be found in Ref.~\cite{Chajewski2023S}.

\textit{\textbf {Crystal structure and chemical composition characterization}}

The x-ray diffraction measurement in combination with Rietveld refinement confirmed that so-obtained crystals adopt tetragonal CaBe$_2$Ge$_2$-type crystal structure (space group \textit{P4/nmm}, no. 129) with lattice parameters \textit{a}~=~4.282~\AA~and \textit{c}~=~9.848~\AA, being close to those reported previously \cite{Khim2021S}. The chemical composition of the samples was examined by electron-microprobe analysis with energy-dispersive x-ray spectroscopy. The determined atomic ratio Ce/Rh/As~=~20.6/39.5/39.9 almost perfectly corresponds to nominal stoichiometry 1:2:2. The quality of samples was checked using Laue diffraction, and the same method was used to orient the crystals. 

\textit{\textbf {Specific heat measurements}}

Specific heat measurements were performed down to 80 mK using both the standard time-relaxation method \cite{Hwang1998S} and long-pulse technique (see e.g. ref. \cite{Scheie2018S}) implemented in a commercial Quantum Design PPMS (Physical Property Measurement System) platform equipped with a $^3$He-$^4$He dilution refrigerator.

\section{Zero-field specific heat data analysis}\label{secSM1}
\vspace{10mm}

\begin{figure}[htbp]%
	\centering
	\includegraphics[width=0.7\textwidth]{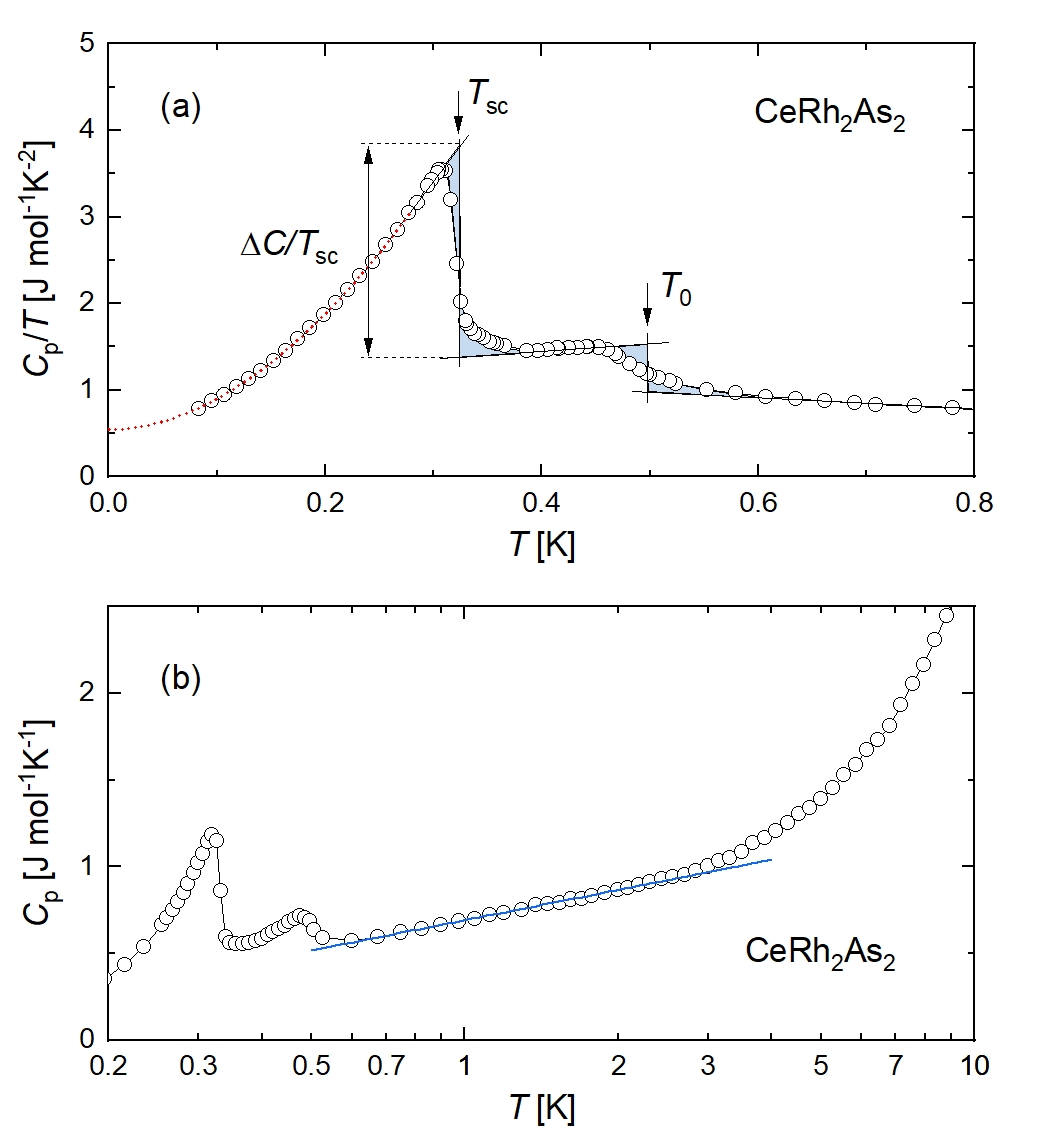}
	\caption{\textbf{Zero-field specific heat of CeRh$_2$As$_2$.} (a) Low-temperature specific heat data of CeRh$_2$As$_2$ with marked transition temperatures, superconducting specific heat jump $\Delta C$ and equal-entropy constructions (shaded areas). Straight lines are the guides for the eye. The red dotted curve is an extrapolation of the lowest-temperature data down to 0 K (see details in the text). (b) Zero-field specific heat data measured in extended temperature range. Blue straight line represents the logarithmic temperature dependence $A\ln(T/T^*)$. } \label{figS1}
\end{figure}

Fig.~\ref{figS1}a shows the low-temperature zero-field specific heat data of CeRh$_2$As$_2$. There is presented a method of determination of the transition temperatures $T_0$ and $T_{\rm sc}$ with the use of the equal-entropy analysis. Based on this construction, we determined the value of the superconducting specific heat jump, $\Delta C$ at $T_{\rm sc}$, which allowed us to calculate the ratio $\Delta C / C\big|_{T_{\rm sc}} = 1.78$. Recently, Semeniuk \textit{et al.} \cite{Semeniuk2023S} pointed to the increase of the height of the superconducting specific heat jump $\Delta C / C\big|_{T_{\rm sc}}$ to about 1.3, compared to around~1 determined in the original paper by Khim \textit{et al}. \cite{Khim2021S}, as one of the indications of higher quality of the crystals. Since the same parameter determined for our single crystals is even higher and reaches a value of about 1.8, we suppose that our samples are of even better quality.

Moreover, we found, that below about 0.28~K, i.e. in the superconducting state, the specific heat data of the compound can be well described by the relation $C_{\rm p}/T = a_0 + b_nT^n$, with parameters $a_0= 0.54$ Jmol$^{-1}$K$^{-2}$, $b_n = 28.2$ Jmol$^{-1}$K$^{-(2+n)}$, and $n = 1.89$. The fit of the aforementioned formula to the experimental data, extrapolated down to 0~K, is presented in Fig.~\ref{figS1} as a red dotted curve. We assume that the parameter $a_0$ is a good approximation of the Sommerfeld coefficient $\gamma = C_{\rm p}/T$ for $T \rightarrow 0$. Its value is significantly lower than about 0.7~Jmol$^{-1}$K$^{-2}$ and 1.2~Jmol$^{-1}$K$^{-2}$ from recent \cite{Semeniuk2023S} and original \cite{Khim2021S} reports, respectively, even though in our analysis (contrary to previous ones) no subtraction of the nuclear contribution to the specific heat has been performed. This probably signals significantly reduced disorder in our single crystals.

The zero field specific heat data measured in extended temperature range is presented in Fig.~\ref{figS1}b. As can be seen, it exhibits the $A\ln(T/T^*)$ temperature dependence (see the blue straight line) in the temperatures up to about 3~K.

\section{Zero-field electrical resistivity data}\label{secSM2}

\begin{figure}[htp]%
	\centering
	\includegraphics[width=0.6\textwidth]{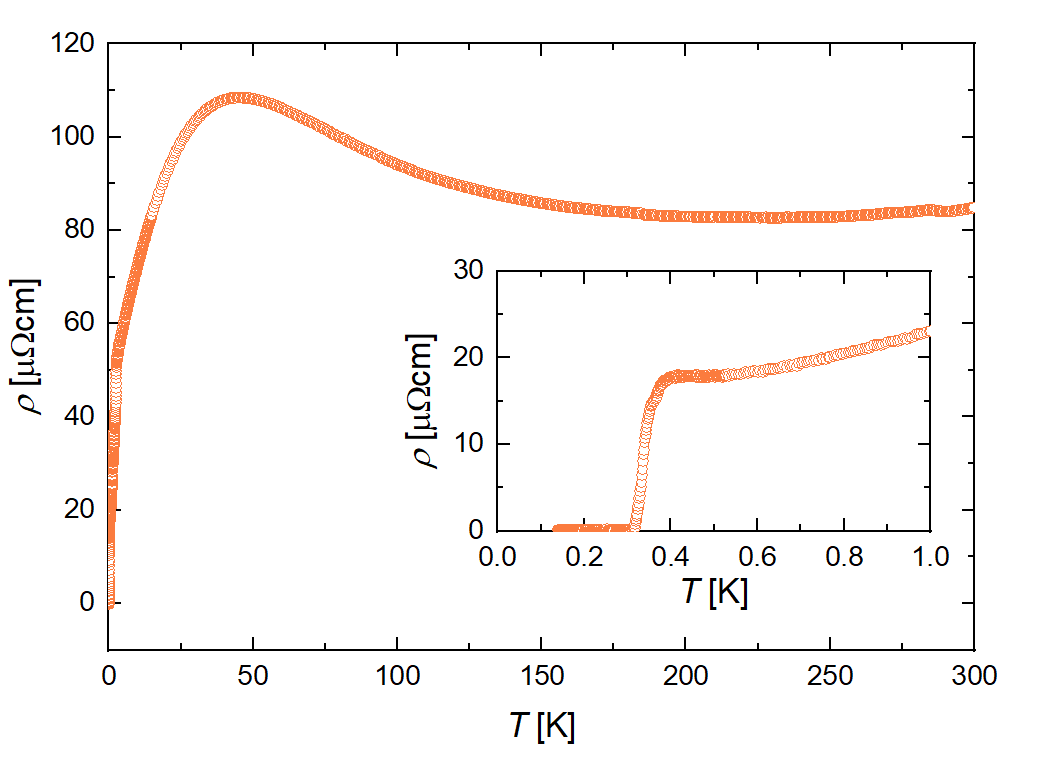}
	\caption{\textbf{Zero-field electrical resistivity of CeRh$_2$As$_2$.} Temperature dependence of the electrical resistivity of CeRh$_2$As$_2$ measured in a zero magnetic field, displayed in a broad temperature range (main panel) and in the lowest temperatures (inset).} \label{figS2}
\end{figure}

Fig.~\ref{figS2} presents the zero field temperature dependence of the electrical resistivity of CeRh$_2$As$_2$. At room temperature $\rho$ takes a value of 84.8 $\mu \Omega cm$, while at low temperatures, it reaches a value of $\rho_0 = 17.8~ \mu \Omega cm$ at 0.5~K. Both higher residual resistivity ratio, defined as $RRR = \rho(300 K)/\rho (0.5 K)\approx 4.8$, compared to 2.8 from the recent article, and lower $\rho_0 \approx 18~\mu \Omega cm$, being about twice lower than the value reported by Semeniuk \textit{et al} \cite{Semeniuk2023S} corroborate the higher quality of our single crystals.

\vspace{10mm}

\section{Evolution of the first order phase transition in the specific heat data}\label{secSM2}

\begin{figure*}[htbp]%
	\centering
	\includegraphics[width=1\textwidth]{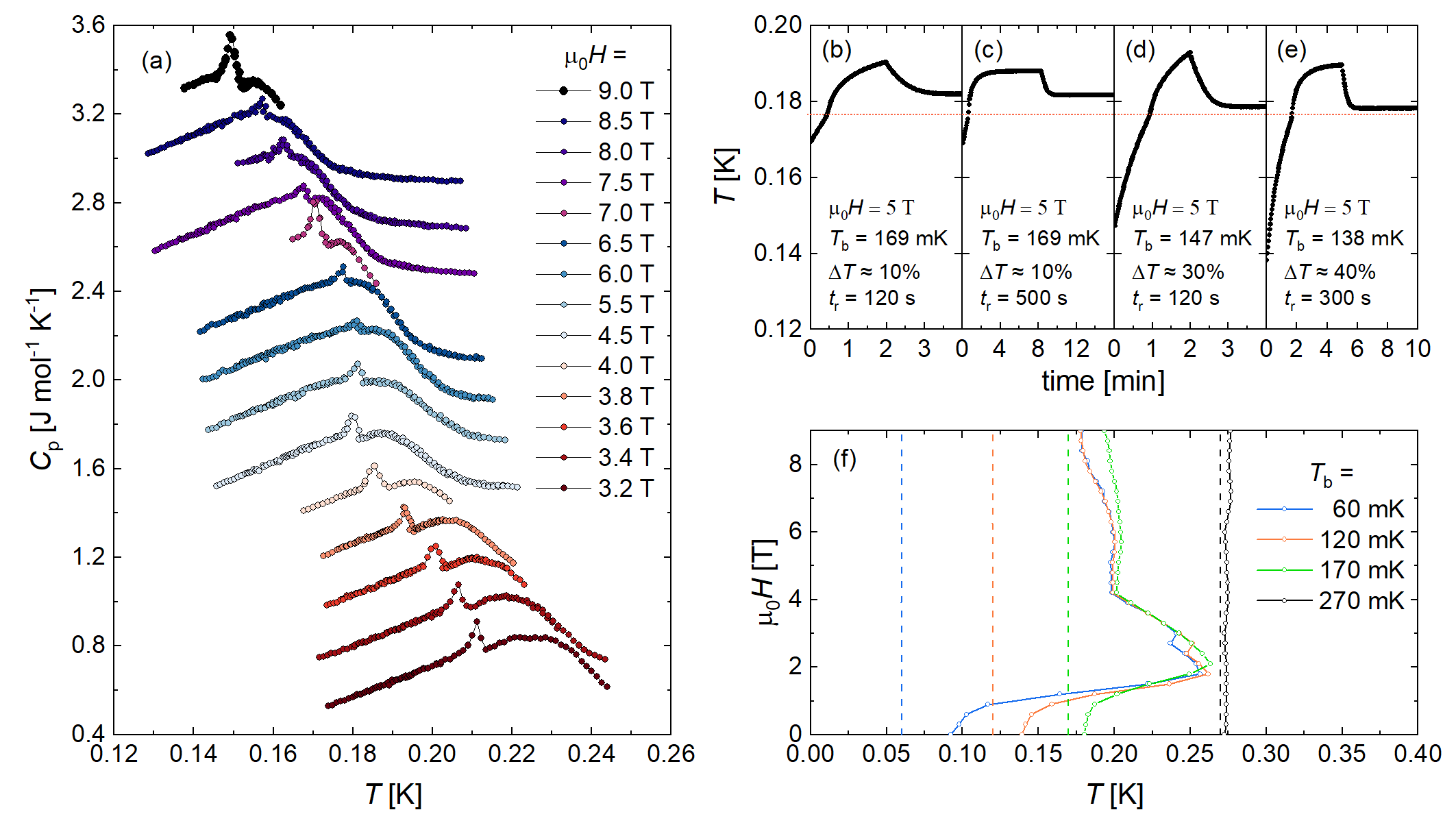}
	\caption{\textbf{Evolution of the first-order phase transition with a magnetic field.} (a) First-order phase transition detected in the specific data collected with the use of the long heat pulse method and standard dual-slope analysis. For the sake of clarity, the consecutive curves were shifted upwards by 0.2~J~mol$^{-1}$~K$^{-1}$.  (b-e) Temperature relaxation profiles measured in a magnetic field of 5~T with various sets of parameters: base temperature of the dilution refrigerator insert ($T_{\rm b}$), temperature rise ($\Delta T$), and relaxation measurement time ($t_{\rm r}$). The horizontal red dashed line marks the first-order phase transition temperature ($T_{\rm 1st}$) derived from the heating curves. (f) The readings of the thermometer attached to the sample platform of the heat capacity puck, measured for various magnetic fields applied. Different curves were collected for a few different base temperatures ($T_{\rm b}$) of the dilution refrigerator insert (the corresponding base temperatures were marked by vertical dashed lines with the same colors). } \label{figS3}
\end{figure*}

In order to more precisely track the evolution of the first-order phase transition, we performed the specific heat measurements with the use of the long heat pulse method near the critical temperature $T_{\rm sc}$ in additional magnetic fields. Examining the obtained data we found that the first-order phase transition is clearly visible (see Fig.~\ref{figS3}a) even with the use of the dual-slope analysis method, which is basically not suitable for the detection of first-order phase transitions. However, compared to the results obtained by the single-slope analysis (as presented in Fig.~1 and Fig.~2 in the main text), the amplitude of the first-order transition anomaly is much smaller in this case. Contrarily, we were not able to detect the presence of the first-order phase transition at $T_{\rm 1st}$ in the specific heat data collected with typically used time-relaxation (semi-adiabatic) technique and analyzed by standard dual-slope method.

Interestingly, in our numerous measurements, we found that if no proper thermalization of the sample platform is ensured and/or low cooling power of the dilution refrigerator is used during the measurement, such a huge specific heat anomaly at $T_{\rm 1st}$ plays a role of a thermal barrier for which a significant amount of heat has to be delivered to or collected from the sample to cross it through. Due to the construction of the specific heat calorimeter puck heating the sample is relatively easy because the heater is directly attached to the sample platform and hence, the heat transfer from the heater to the sample is fast. On the contrary, for cooling the heat has to be dissipated from the sample platform to the mounting stage of the dilution refrigerator, and it is possible only through the wires and quartz fibers supporting the sample platform. Therefore, this process is much slower. We decided to use the above-mentioned feature to determine how the first-order phase transition temperature $T_{\rm 1st}$ changes with magnetic fields. Because for each measurement several different parameters (base temperature of dilution refrigerator insert ($T_{\rm b}$), temperature rise ($\Delta T$), and relaxation measurement time ($t_{\rm r}$)) can be adjusted, we performed a number of test runs on a 0.4~mg crystal in order to define how these factors influence the measured transition temperature. The exemplary heating-relaxation temperature profiles obtained in the magnetic field of 5~T for various sets of parameters are presented in Figs.~\ref{figS3}b-d. As can be noticed, upon heating the sample, the first-order phase transition, which is observed as an abrupt slope change in the heating curve, seems to occur at the same temperature (marked by the horizontal red dashed line in Figs.~\ref{figS3}b-d), independently of the parameters used. However, because during the heating the crossing $T_{\rm 1st}$ happens very fast, the determination of this temperature is rather imprecise. On the other hand, during the cooling, it can be determined very precisely by conducting the measurement for a sufficiently long time and assuming that the temperature of the relaxation curve saturation reflects $T_{\rm 1st}$. In our studies, we found that for the used  sample its temperature becomes stable after about two minutes of cooling. Further increase of $t_{\rm r}$ does not result in a significant change of the saturation temperature (compare Figs.~\ref{figS3}b and \ref{figS3}c). Furthermore, we did not observe any clear correlation between the used temperature rise $\Delta T$ and measured saturation temperature $T_{\rm 1st}$. In turn, there is an apparent relationship between $T_{\rm 1st}$ and the base temperature of the dilution refrigerator. In general, for lower $T_{\rm b}$, we measured lower $T_{\rm 1st}$ (see Figs.~\ref{figS3}c-e). However, for sufficiently low base temperatures and in magnetic fields higher than about 3~T, the change of $T_{\rm b}$ does not influence visibly the measured saturation temperature of the relaxation curve.

Fig.~\ref{figS3}f presents the readings of the sample platform thermometer ($T_{\rm sp}$) measured for several different base temperatures of  the dilution refrigerator and in a wide range of magnetic fields. To ensure enough time for the sample temperature to stabilize and to avoid heating the sample due to the magnetic field change, each data point was collected several minutes after the magnetic field had been set. As can be seen, for $T_{\rm b} = 60$~mK, 120~mK, and 170~mK (blue, orange, and green curves, respectively), the readings of the sample platform thermometer  differ significantly from the corresponding base temperatures (blue, orange, and green dashed lines, respectively) in the whole studied range of magnetic fields. In turn, for $T_{\rm b} = 270$~mK (black curve) no significant differences between the sample platform and dilution refrigerator temperatures are observed. It is due to the fact, that in this case, the first-order phase transition occurs at temperatures lower than $T_{\rm b}$ and thus, no thermal barrier has to be crossed to bring the sample temperature to about 270~mK. In other cases, the situation is reversed and the barrier prevents the sample from reaching the base temperature. It is also worth noting that in the progression of the first three curves, two different regimes may be distinguished. In the first, low field one, measured $T_{\rm sp}$ clearly depends on the base temperature of the measurement. For lower $T_{\rm b}$, the observed readings of the sample platform thermometer are also lower. This means that in the low field range, the position of the first-order phase transition strongly depends on the cooling speed, possibly due to the supercooling of the system. One can also note that in low fields (up to about 2~T), $T_{\rm sp}$ increases with increasing magnetic field. Similar behavior was previously observed in specific heat data of Ce$_2$RhIn$_8$, where initially broad, and not indicating a first-order phase transition, zero-field anomaly, with increasing magnetic field becomes sharper and shifts toward higher temperatures \cite{Lengyel2008S}. This similarity is in favor of the AFM ordering scenario in CeRh$_2$As$_2$. In the second regime, above about 3~T, the curves measured for $T_{\rm b} = 60$~mK and 120~mK superimpose onto each other perfectly, marking the position of the first-order phase transition. The curve measured for $T_{\rm b} = 170$~mK slightly deviates from them towards higher temperatures in magnetic fields higher than about 4~T, which is most likely due to the much smaller difference between $T_{\rm b}$ and $T_{\rm 1st}$.


%

\end{document}